\makeatletter
\pdfoutput=1 
\makeatother
\documentclass[11pt,american,english]{article}
\usepackage[latin9]{inputenc}
\usepackage{float}
\usepackage{algorithm2e}
\usepackage{amsmath}
\usepackage{amssymb}
\usepackage{graphicx}

\makeatletter
\pdfoutput=1
\usepackage{latexsym}
\usepackage[displaymath]{lineno}
\usepackage{graphicx}

\usepackage{newfloat}
\DeclareFloatingEnvironment[placement={!ht},name=List]{mylist}

\makeatother

\usepackage{babel}
\begin{document}
\title{On the importance of system-view centric validation for the design
and operation of a crypto-based digital economy}
\author{Alexander Poddey\thanks{Bosch Center for Artificial Intelligence, Robert Bosch GmbH, Corporate Research Campus, 71272 Renningen, Germany,  alexander.poddey@de.bosch.com}\and Nik Scharmann\thanks{Project Director Economy of Things, Robert Bosch GmbH, Corporate Research Campus, 71272 Renningen, Germany}  }
\date{\the\year}

\maketitle
\begin{abstract}Ubiquitous connectivity, networked computation, open
technologies and advances in intelligent web approaches (like semantic
web, distributed databases, and intelligent applications) enable the
third-generation web, Web3.0. Not least due to advancements in crypto
technologies adoption in real world applications, it is now possible
to convert the internet of things (IoT) into an economy of things
(EoT), which basically refers to a heterogeneous digital economy of
everything. Based on the realization of the deficiencies of Web2.0
and the loss of confidence in a central intermediaries based economy,
a societal desire to overcome these deficiencies has emerged. The
so-called social Web2.0 in fact is dominated by a few quasi monopolistic,
central platforms incorporating most of the utility arising e.g. by
collecting large amounts of user specific data to feed machine learning
algorithms and provide proprietary services based on this. The banking
crisis on the other hand demonstrated that the broad society is at
risk if these central intermediaries fail. Cryto-technological advancements
led to the development of technology, enabling a secure digital economy
based on peer to peer trust-less networks, thereby effectively eliminating
the need for central intermediaries (which in history regularly demonstrated
their fallibility).

In the beginning, the crypto movement has been enthusiastic about
the ability to encode all necessities for a digital economy in algorithms
and code. However, in the mean time, insight has gained ground that
on the one hand, designing the \emph{EoT} is a quite complex endeavor
and the resulting system might fail to achieve the targeted ideals,
even when no intermediaries are present. We emphasize this position
by arguing that the \emph{EoT }in fact corresponds to a \emph{complex
system operated in open contexts}, for which a sophisticated design
approach and a set of measures are inevitable, in order to be able
to achieve a valid design and operation.

In addition, we argue that the high level guiding principles necessary
for a valid and effective \emph{EoT} align well with the ideals of
the democratized Web3.0 movement, and are no naive rapture, but rather
- adequate adoption provided - can lead to socioeconomic efficiency.
Over and above, we postulate that in this kind of \emph{EoT,} commercial
companies not only play a role, but can benefit alongside the society.
In more game theoretical jargon, one could state that it is rational
for the society \emph{and} enterprises to undertake the effort to
establish a valid \emph{EoT }based on these ideals\emph{.} 

\end{abstract}

\section{Introduction}

Ubiquitous connectivity, networked computation, open technologies
and advances in intelligent web approaches (like semantic web, distributed
databases, and intelligent applications) enable the third-generation
web, \textbf{\emph{Web3.0}}. Not least due to advancements in crypto
technologies adoption in real world applications (initiated by Satoshi
Nakamoto and bitcoin \cite{Nakamoto2008}) it is now possible to convert
the internet of things (\textbf{\emph{IoT}}) into an economy of things
(\textbf{\emph{EoT}}), which basically refers to a heterogeneous digital
economy of everything \cite{Pureswaran.2015b}(see section \ref{subsec:EoT,-EoT-functions}
for details).

Based on the realization of the deficiencies of Web2.0 and the loss
of confidence in a central intermediaries based economy \cite{Finley.2016},
partly due to the banking crisis, a societal desire to overcome these
deficiencies has emerged \cite{Ecorys.2017}. The so-called social
Web2.0 in fact is dominated by a few quasi monopolistic, central platforms
incorporating most of the utility arising e.g. by collecting large
amounts of user specific data to feed machine learning algorithms
and provide proprietary services based on this. The banking crisis
on the other hand demonstrated that the broad society is at risk if
these central intermediaries fail. The hallmark paper of Satoshi Nakamoto
\cite{Nakamoto2008} and the thereby kicked of cryto-technological
advancements led to the development of technology, enabling a secure
digital economy based on peer to peer trust-less networks, thereby
effectively eliminating the need for central intermediaries (which
in history regularly demonstrated their fallibility).

In the beginning, the crypto movement has been enthusiastic about
the ability to encode all necessities for a digital economy in algorithms
and code. However, in the mean time, insight has gained ground that
on the one hand, designing the \emph{EoT} is a quite complex endeavor
and the resulting system might fail to achieve the targeted ideals,
even when no intermediaries are present. On the other hand, it becomes
more and more obvious that the algorithmic part needs to be accompanied
by extra-algorithmic measures, such as social mechanisms for self-
and societal-governance (see section \ref{sec:EOT:-a-complex} for
a detailed discussion).

We emphasize this position by arguing that the \emph{EoT }in fact
corresponds to a \emph{complex system operated in open contexts},
for which a sophisticated design approach and a set of measures are
inevitable, in order to be able to achieve a valid design and operation,
as argued in our previous work \cite{Poddey2019} In this respect
we extend the design requirements posed upon intelligent economic
networks \cite{Heylighen.2015}. We illustrate the relation of the
sys$^{2}$val's basic methodological aspects to the \emph{EoT} design
process in section \ref{sec:EOT:-a-complex}.

In addition, we argue that the high level guiding principles necessary
for a valid and effective \emph{EoT} align well with the ideals of
the democratized Web3.0 movement, and are no naive rapture, but rather
- adequate adoption provided - can lead to socioeconomic efficiency.
Over and above, we postulate that in this kind of \emph{EoT,} commercial
companies not only play a role, but can benefit alongside the society.
In more game theoretical jargon, one could state that it is rational
for the society \emph{and} enterprises to undertake the effort to
establish a valid \emph{EoT }based on these ideals\emph{.} 

Throughout the text, we use \textbf{\emph{boldface}} characters for
term definitions, whereas \emph{italic} characters indicate the use
of already defined terms elsewhere within the document and \emph{italic}$^{*}$references
terms defined in \cite{Poddey2019}. 

\section{\emph{EOT/dEoE:} a complex system operated in open context\label{sec:EOT:-a-complex}}

\subsection{\emph{EoT}, \emph{dEoE}, functions, agents and general digital intelligence\label{subsec:EoT,-EoT-functions}}

The term economy of things (\emph{EoT}) evolved from internet of things
(\emph{IoT}). Internet of things refers to the fact that nowadays,
due to ubiquitous connectivity, not only humans connect via the web.
It is also possible to build networks of things like sensors, fridges,
cars - so called \textbf{\emph{IoT devices}}. However, connecting
everything with everything by itself is of no use. The connected entities
need to be able to interact in ways comparable to established economic
mechanisms such as search and find, negotiation, payment, settlement,
building trust etc. in order to make use of the connectivity. The
\emph{IoT} therefore needs to be converted into an economy of things.
However, although broadly used, the trailing part 'of things' is misleading.
In fact, what is meant by \emph{EoT} is a digital economy of everything
(\textbf{\emph{dEoE}}) - a heterogeneous mix of e.g. small \emph{IoT}
devices, more powerful digital entities like machine learning based
services running in the cloud and humans, interacting seamlessly. 

In Web2.0, functionality was basically centralized. The functions
of a service could in fact be split into several modules, but there
typically is a central point of service providing access. In Web3.0
even smaller modules can be incorporated as individual entities, connecting
and interacting with others on their own behalf. These entities are
usually called \textbf{\emph{agents}}. Web3.0 therefore can be understood
as a \textbf{\emph{multi-agent}} system in which functionality - or
in a more general form the \textbf{\emph{capability}} to achieve a
goal - emerges from interaction of fragmentary contributions \cite{Wooldridge2009,Wooldridge2013}.
The capability therefore is no longer embodied in a monolithic entity,
but distributed across a network of agents, each embedding only a
part of the necessary modules. Over and above, individual agents contributions
to the network are not guaranteed regarding quality of service and
reliability. Agents providing new fragmentary contributions might
appear, others disappear. The \emph{dEoE} therefore is a prime example
of a complex open context system. See \cite{Poddey2019} for a detailed
definition and discussion of the challenges related to the valid design
of such systems.

Due to the fragmentation into (sub-)modules and the possibility to
flexibly and dynamically combine these to new and more complex compositions
\emph{dEoE} will not only lead to an ever growing capability of the
network. Over and above, capabilities will emerge, no one of the designers
of the individual agents has been thought of beforehand (\emph{emergent
behavior}$^{*}$). The ability of the network to evolve and solve
unforeseen tasks might be regarded as a form of intelligence \cite{ShaneLegg.2007}.
This is a natural effect in complex, modular systems and has nothing
artificial. We therefore refer to the capability of the \emph{dEoE
}as \textbf{\emph{digital intelligence (DI)}}. In contrast to the
nowadays mainly monolithic machine learning systems, which are extensively
hyped as artificial intelligence, on a closer look however turn out
to be at most comparable to quite narrow intelligence, \emph{DI} on
the one hand continuously evolves by itself and might therefore in
the future lead to what we call \textbf{\emph{general digital intelligence
(gDI)}}. In addition, \emph{dEoE} seamlessly integrates heterogeneous
agents (be they representatives of humans or autonomous digital entities)
and therefore by its basic construction allows to leverage the synergistic
human machine potential. We therefore argue that \emph{(g)DI} will
be beyond \emph{(g)AI}.

Based on this insight, it becomes clear that the huge potential of
\emph{Web3.0} \& \emph{dEoE }requires a responsible\emph{ holistic$^{*}$
}handling. Due diligence is necessary in design and operation of the
total system, specifically also covering the emerging capabilities
and effects not even known at individual components design time, just
as argued in \cite{Poddey2019}. Due to the main utility arising from
this emergence, it is by no means sufficient to seemingly perfect
design and validate on individual modules or functions basis, e.g.
stable coin and payment service. It is rather highly essential to
apply valid design and operation approaches on a \emph{holistic} basis
to the total system. Not least to prevent a comparable outcome to
Web2.0, where the utility and power aggregated in the hands of a few
quasi platform monopolists, taking advantage of it to the detriment
of smaller competitors and society. 

Not approaching the \emph{dEoE} in a \emph{holistic} manner might
allow the quasi platform monopolists to extend their power and dominance
even further. As an example, think of a perfectly orchestrated open
source based initiative, seemingly well designed and governed, however
only on the individual function level like e.g. payment. Combine this
function with Web2.0 services, such as e.g. social networks, allowing
the operator of the platform to then extract crypto-payment related
information and combine it with the rest of the already available
data from the platform users. The effect of such an approach from
a total system level view is by no means in accordance with the ideals
of the democratized Web3.0 movement and detrimental to all other
participants of the \emph{dEoE}. See section \ref{sec:The-high-level}
for a detailed discussion.

\subsection{Recap of sys2val and relation to \emph{dEoE\label{subsec:Recap-of-sys2val}}}

\emph{dEoE} is a prime example of a complex open context system, as
argued in the previous section. Due to the\emph{ implicit infinite
complexity}$^{*}$, these systems can not validly be designed, maintained
or operated in an ad-hoc manner. Instead, more rigorous (systematic
and holistic) approaches are required. See \cite{Poddey2019} for
a detailed discussion and the introducion of sys$^{2}$val, a holistic,
systematic and system-based view centric approach providing such strategies.
State of the art for complex closed context systems engineering is
the holistic approach based on system view, as established by Rasmussen
\cite{Rasmussen1994} and Leveson \cite{Leveson2011}, with a good
track record e.g. in aviation, military and aeronautics. Sys$^{2}$val
is based on this strong fundament and extends it to enhance traceability,
completeness- \& validation and applicability for open context systems.

In a nutshell, complex open context systems can never be designed,
validated and operated in a formal complete manner. In contrast to
closed context systems, unpredictability needs to be embraced and
dealt with on a systematic basis. The basic ingredients are
\begin{itemize}
\item get clear about what should be achieved, based on which ideals and
high level guiding principles (\emph{high level goal}$^{*}$, \emph{aimed
purpose}$^{*}$)
\item iteratively concretize the functional and non-functional requirements
while being aware of related assumptions taken (\emph{valid deductive
step}$^{*}$)
\item do as much as you can based on system understanding (\emph{validation
challenge}$^{*}$, \emph{holistic cycle}$^{*}$)
\item accept that an a priori complete understanding of later system dynamics
and effects will not be possible and therefore prepare well for the
unexpected (\emph{validation challenge}$^{*}$, \emph{holistic cycle}$^{*}$)
\end{itemize}
This translates in the requirement for the design of an antifragile
system \cite{Taleb2012,Taleb2013,Taleb2014a}, implementing continuous
improvement based on observation (\emph{monitors}$^{*}$). In order
to be able to do so, it is necessary to apply prepared and continuously
adapted \emph{recovery$^{*}$} and \emph{degradation strategies$^{*}$}
(survive the unexpected), feedback the observed to extend system understanding
(learn from it) leading to ongoing improvement (get better). This
is a continuous process based on a \emph{holistic cycle}$^{*}$.

Regarding societal complex systems, which \emph{dEoE} is a representative
of, the design of four mutually dependent basic aspects need to be
addressed with due diligence, following the outlined approach. Namely: 

\begin{mylist}\begin{itemize} 
\item algorithmic mechanisms and protocols
\item algorithmic mechanisms based self-governance
\item social mechanisms based self-govenance
\item societal governance
\end{itemize} \caption{The four mutually dependent aspects which necessarily need to be addressed in the design of complex societal systems.}\end{mylist}

The first two aspects  are, in principle, taken serious and addressed
via game theory based mechanism design. Having said this, it needs
to be added that it holds for the implementation of certain goals
like e.g. decentralization and privacy. However, there is no widely
discussed and accepted set of goals such systems should implement.
This set obviously should include much more than the aforementioned
and is an involved matter on its own. 

Self-governance of Web3.0 projects, especially in the crypto field,
is nowadays mostly following an informal, technocratic model \cite{Yeung2019,Srinivasan2017}.
This is not necessarily the best approach for systems with a large
inherent potential to affect the broad society. It is therefore more
and more frequently criticized. 

And last but not least, due to the rapid advances in the field, inter-
and national regulation - mainly responsible to establish a societal
governance - lacks well behind (see e.g. the discussion in \cite{Hamelink2003}).
However, it is important for projects potentially having an impact
on the broad society, to align with inter- and national regulation
in order to close the gap. Otherwise, disagreement might arise, as
the discussion around the Libra project illustrates \cite{Waters2019}. 

In conclusion, a widely discussed and accepted set of design goals
for the \emph{dEOE }should to be established\emph{.} In addition,
it needs to be well understood that these goals can not efficiently
be implemented based solely on algorithmic mechanisms. It rater requires
a well balanced combination of all the four mentioned aspects. We
therefore propose and discuss a \emph{high level goal}$^{*}$, its
individual aspects and derived \emph{unacceptable losses$^{*}$} and
\emph{hazardous system states$^{*}$} in the sys$^{2}$val sense in
section \ref{sec:The-high-level}, to stimulate a broader discussion.

These high level ingredients then allow to derive a well balanced
combination of measures to the achievement of a valid and efficient
design and operation of \emph{dEoE}. 

Before detailing out the concrete high level goal, an overview of
its role and the relation to the four aspects of listing 1 will be
given in the following section.

\section{The role of the high level guiding principles in the constitutional
process of \emph{dEoE}}

\subsection{Introduction}

The holistic, systematic and system-based view centric approach\emph{
}to the design of complex open context systems, in which emergence
plays an important role, has been discussed in the foregoing sections
and \cite{Poddey2019}. In a nutshell, high level guiding principles
are used during derivation of the iteratively concretized design and
ongoing improvement of the targeted functions of the (total) system
such, that the dynamically evolving system, more specifically its
\emph{emergent behavior}$^{*}$, complies with these principles. To
this end, monitoring of underlying assumptions, prepared recovery
and degradation strategies as well as mechanisms for continuous improvement
and adaption (\emph{valid deductive steps}$^{*}$) are necessary.
The development goal is to achieve a well balanced and efficiently
operating combination of the four aspects given in listing 1 in the
previous section, on all levels of concretisation (see also \emph{development
goal}$^{*}$). 

\begin{figure}[H]
\begin{centering}
\includegraphics[viewport=0bp 80bp 800bp 550bp,clip,width=0.8\textwidth]{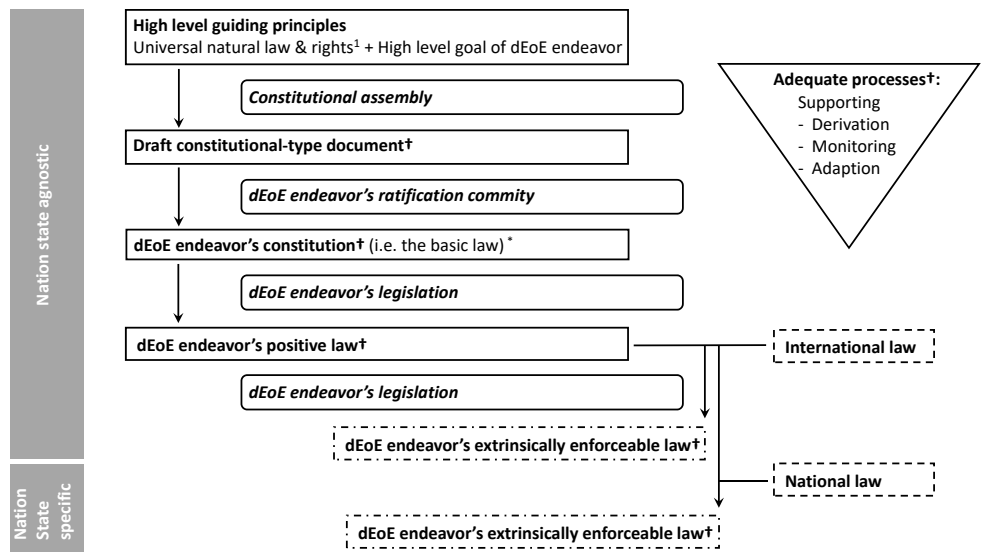}
\par\end{centering}
\caption{\label{fig:const}Overview of the proposed process from the high level
guiding principles to the nation specific enforceable law. See the
text for a detailed discussion.}
\end{figure}

Figure \ref{fig:const} provides an overview of a constitutional process
we suggest for a \emph{dEOE} endeavor. This process will be discussed
in the following.

We deliberately choose the abstract term 'endeavor' to indicate that
the process is independent of the type of initiative. It may be applied
to the formation of a specific legal entity for the operation of certain
aspects of the \emph{dEoE, }the formation of domain specific alliances
etc. However, as discussed above, the high level goal always relates
to the emergent effects of the total system.

The necessary elements of an adequate sys$^{2}$val based approach
are subsumed in the triangular box to the right. The pervasive character
of the application is indicated by the $\dagger$ added to all elements
representing results of a \emph{deductive step}$^{*}$. 

\subsection{From high level guiding principles to the basic law}

The most abstract level is formed by the \textbf{\emph{universal natural
law and rights}}. Universal thereby refers to principles and concepts
to be most legitimate, not subject to relativism. Universal rights
exist independently of human understanding and of the \textbf{\emph{positive
law}} - and therefore are widely accepted. The term \emph{positive
law}, from\textquoteleft to posit\textquoteright , in contrast, relates
to the law of a given state, political order, legislature or society.
The natural law and rights for example include universal human rights
\cite{UDHR1948} and international covenant on civil and political
rights - ICCPR \cite{UN1966}, from which i.a. the right to privacy
in information technology can be derived (specifically Article 12
UDHR and 17 of ICCPR), see e.g. \cite{Hamelink2003}. We propose to
augment the natural law and rights by the aspects of the high level
goal of \emph{dEoE} endeavors, discussed in section \ref{sec:The-high-level}.
Together, these high level guiding principles form the input to the
constitutional assembly.

Aspects of the type 'guiding principles, rights and laws' are indicated
by rectangular boxes with sharp corners. Specific organs, typically
a subset of all concerned entities (indicated by boxes with rounded
corners), take \emph{deductive steps$^{*}$} in order to derive a
concretization of the more abstract level to a more endeavor-specific
interpretation and elaboration. The formation of these organs itself
needs to obey the high level guiding principles; they need to be arguably
adequate (i.e. \emph{valid}) with respect to the formation (e.g. who
can be member of the committee, how are these members selected, etc.)
and exertion processes. 

The constitutional assembly prepares a draft constitutional-type document,
for which a ratification process, typically involving a broader set
of community members, is conducted. This combination, both, ensures
efficiency and broad acceptance of the resulting \textbf{\emph{dEoE
endeavors constitution}}. The constitution is also referred to as
the\textbf{\emph{ basic law}}. We suggest to apply the doctrine of
separation of powers, keeping the three aspects of governance, namely
legislative, judicial and executive separate and thereby establishing
a system of mutual checks and balances\footnote{Note that figure X only shows the organs related to the deduction
of rights and laws. It does not include further necessary organs,
e.g. judicial and executive institutions. }. As will be discussed in section \ref{sec:The-high-level}, the requirement
for separation of powers directly follows from the high level guiding
principles and is related to the aspect of socioeconomic efficiency.

\subsection{Positive law}

The \emph{dEoE} endeavor's legislature derives the \emph{dEoE} endeavor's
\emph{positive law}. Basically, the \emph{positive law} is a set of
provisions and rules for application in day to day business. All aspects
of listing 1 are covered by these rules. An example for algorithmic
mechanisms and protocols is the well known longest chain rule. Contract-encoded
voting, and the process of opinion making preceding a voting are a
combination of algorithmic (the voting contract) and social mechanisms
(the opinion making). An example for the latter would be a rule which
prescribes that, for certain kinds of voting, the pros and cons of
the alternative choices need to be documented and accessible for a
certain time frame before the voting to all who hold a voting right.
In addition, it typically needs to be regulated who has voting rights,
the weighting of the individual votes as well as by whom and under
which condition alternative choices can be added to the voting.

It is important to note that the \emph{dEoE} endeavor's positive law
is not necessarily \emph{extrinsically enforceable }(i.e. enforceable
from outside the endeavor)\emph{.}The positive law necessarily needs
to respect (i.e. not violating) international law - otherwise operation
of the endeavor might be \emph{extrinsically} shut down, based on
this. However, some aspects not covered by international law can only
be enforced \emph{intrinsically} (from within the endeavor). An example
for an\emph{ intrinsically enforcement} is, again, the longest chain
rule. The system's incentives are - based on game theory - actually
designed such, that it is enforced automatically. It is not intended
and in fact impossible to take node operators not respecting the longest
chain rule to court for not doing so.

This illustrates, that there are several different organs necessary
for implementation of checks and balances. The set of organs is a
complex conglomerate with \emph{dEoE} endeavor intrinsic and -extrinsic
contributions, together adequately covering the four aspects given
in listing 1 and establishing separation of powers, e.g. by splitting
legislature, judicial and executive.

\subsection{Alignment with (inter-) national law}

As discussed above, the endeavors \emph{positive law} needs to be
aligned with international law. At least not violating it. The endeavors
legislature also determines which parts of the constitution should
be covered in the basic law such, that it is explicitly \emph{extrinsically
enforceable} by international, or even nation state specific law.
This, on the one hand serves credibility and on the other hand might
even be necessary in order to be able operate or provide service in
certain states and regions. It is however important to regulate as
much as possible in a nation state agnostic form, for the endeavor
not becoming a plaything or fall victim of individual nation states.
This separation is indicated in the lower part of the figure.

As stated above, inter- and national regulation (indicated by dashed-line
boxes) - mainly responsible to establish a societal governance - lacks
well behind the technical development. The high level guiding principles
and sys$^{2}$val based approach to the constitutional process supports
the establishment of an effective societal governance, as it works
out the relevant aspects which necessarily need to be extrinsically
enforceable and therefore should be covered by (inter-) national regulations.

\subsection{Evolution of rule sets}

The necessary elements of an adequate sys$^{2}$val based approach,
subsumed in the (black box) to the right, also include mechanisms
for discourse, decision about and, where appropriate, adaption of
the related levels guiding principles, rights and laws. The implementation
may be inspired e.g. by Hart's concept of law \cite{Hart2012}, as
proposed by \cite{Yeung2019}. It is important to note that the permanence
of the rule set increases with the level of abstraction. The High
level guiding principles are mainly fix. Constitution needs adaption,
however only on a sound basis and following a detailed examination,
whereas the positive law, and in particular jurisprudence, allows
for faster adaption.

\subsection{How to approach the process}

On first sight, the necessary constitutional process for \emph{dEoE}
endeavors seems complex. In addition, it might be argued that there
is no state of the art yet. However, there is a huge state of the
art e.g. in human societal organization, self governance of open source
projects etc., from which can be drawn. It is therefore not necessary
to reinvent all aspects anew, but to combine understood and established
elements based on the sys$^{2}$val scheme such, that the high level
guiding principles are obeyed. 

As usual with complex systems operated in open contexts, we can not
a priori design a perfectly valid solution right from the start. Therefore,
it is important to establish the processes and measures allowing for
an \emph{antifragile} evolution of the system based on an \emph{ongoing
holistic cycle$^{*}$} (as discussed in section \ref{subsec:Recap-of-sys2val}
and \cite{Poddey2019}, especially section 2 and 4 ibid.). Phrased
differently, sys$^{2}$val provides the basic principles and measures
allowing to establish a lightweight \textbf{\emph{meta governance}},
which then, together with the high level guiding principles, allows
to start an \emph{dEoE} endeavor on a sound basis.  

\section{Guiding principles for the digital economy of everything (\emph{dEoE})\label{sec:The-high-level}}

\subsection{Overview}

In system-view based approaches (e.g. STAMP {[}2{]} and sys$^{2}$val
\cite{Poddey2019} ), the high level goal represents the most abstract
and concise statement of the goal to achieve. All aspects such as
guiding principles, concrete rules and system design are derived from
this by a well established process of iterative concretisation.

The negative effect of non-achievement of aspects of the high level
goal is referred to as \emph{loss}.\emph{ Losses} large enough to
undermine the achievement of the overall goal are referred to as \emph{unacceptable
loss}. A system state possibly leading to unacceptable losses is referred
to as \emph{hazardous system} state and rated as invalid, as it bears
unreasonable risk. 

Section \ref{subsec:Prerequisites} collects some prerequisites for
the formulation and discussion of the high level goal for \emph{dEoE,}
as proposed in section \ref{subsec:The-high-level}. The derived unacceptable
losses and hazardous system states are then given in section \ref{subsec:Unacceptable-losses}.

\subsection{Prerequisites\label{subsec:Prerequisites}}

\subsubsection{Open-endedness}

The adjective 'open-ended' indicates an unrestricted, broad situation
with no fixed limits. In the context of evolution of digital systems,
it relates to the ongoing development of ever complex capabilities
based on emergence in dynamically evolving systems \cite{Banzhaf.2016,TimTaylor.2018}.
As discussed in section \ref{subsec:EoT,-EoT-functions}, due to the
possibility to flexibly and dynamically combine the individual (sub-)modules
of the digital economy to new and more complex compositions, \emph{dEoE}
will not only lead to an ever growing capability of the network. Over
and above, capabilities will emerge, no one of the designers of the
individual agents has been thought of beforehand. By this, an ever
growing utility will be unlocked.

However, open-endedness does not mean fully unrestricted, as will
become clear in the following discussion. In a nutshell, in order
to achieve an efficiently growing capability (and hence utility) of
\emph{dEOE}, a well balanced set of regulations needs to be applied.

\subsubsection{Super modularity}

Super modularity, in short, refers to the popular saying that a combination
of elements provides more utility than the sum of individual elements
utilities. A more formal representation for functions defined over
subsets of a larger set is as follows:

Let $S$ be a finite set. A function $f$ is \textbf{\emph{supermodular}},
if for any $A\subset S$ and $B\subset S$:

\begin{equation}
f(A\cup B)+f(A\cap B)\ge f(A)+f(B)\label{eq:1}
\end{equation}

This is a more strict condition than \textbf{\emph{superadditivity}},
for which 
\begin{equation}
f(A\cup B)\ge f(A)+f(B)\label{eq:2}
\end{equation}
See e.g. \cite{Milgrom1996}. The difference is in the inclusion of
the intersection term $f(A\cap B)$ in eq.\ref{eq:1}. In order to
see the effect of this term, think of two partly overlapping sets.
The resulting overlap from combining these sets ($A\cap B$) may lead
to a negative effect ($f(A\cap B)\le0$). If the combined effect (the
left hand side of eq.\ref{eq:1}) still is larger than the summed
individual effects (right hand side), the combination still pays off.

\subsubsection{Coalition games, coopetition and efficiency}

Supermodularity is related to convex coalition games, and the solution
concept introduced by Shapley \cite{Shapley1953,Shapley1971,Roth1988}.
In coalition games, a set of competitive players may cooperate (on
some aspects) to form a coalition. This might e.g. relate to resource
sharing, sharing of development costs for basic technologies or sharing
of modular capabilities to achieve more complex capabilities, unlocking
a surplus. Put simply, even though there might be some aspects for
which the players are in competition and on which the cooperation
might have a detrimental effect (see the discussion of the intersection
term of eq.\ref{eq:1}), in supermodular systems, cooperation pays
off and everybody taking part benefits. The combination of competition
and cooperation at the same time is referred to as \textbf{\emph{coopetition}}.

Coopetition in supermodular systems generates a surplus. The larger
the set of agents taking part, the larger the total surplus. Therefore,
the goal is to achieve what is called the 'grand coalition'.

A solution to the problem of establishing a stable grand coalition
has e.g. given by Shapely \cite{Shapley1953,Shapley1971,Roth1988}.
It is based on the distribution of the generated surplus according
to the contributions of the individual players. In a nutshell, if
any players benefit from the surplus over-proportionately with respect
to their to contribution, it is beneficial for the other players to
form a smaller coalition excluding them. By this split-up however,
the surplus of the total system drops, which is socioeconomically
undesirable.

Due to the fragmentation of the \emph{dEoE} into (sub-)modules and
the possibility to flexibly and dynamically combine these to new and
more complex compositions, leading not only to an ever growing capability
of the network but also to emerging capabilities no one of the designers
of the individual agents has been thought of beforehand, we argue
the \emph{dEoE} to be supermodular. Therefore, it is important to
establish a \emph{coopetition} based \emph{dEoE} based on adequate
surplus sharing, both, from a viewpoint of most efficient capability
increase towards \emph{gDI}, as well as socioeconomic sense.

As a side note, the problem of adequate distribution of a total surplus
among the players is, on a fundamental basis, also related to the
credit assignment problem of machine learning and AI. In \emph{dEoE,}
both topics comes together.

\subsubsection{Imperfect information and the non-existence of the invisible hand\label{subsec:Imperfect-information-and} }

The discussion of coalition games and supermodularity already indicated
that an efficient generation of surplus and growth of overall capability
will, in general, not necessarily arise automatically. It depends,
i.a., on the involved topic of surplus distribution, related to credit
assignment. The challenge of adequate credit assignment and surplus
distribution lies in the non-locality of effects and hence in the
impossibility to achieve perfect information for rating a certain
contribution or action at a given point in time and system state.
For example taking a certain action in a certain system state, or,
in the coopetition model, contributing a certain element, typically
takes effect in a time-, state- and network participant distributed
fashion. There may even arise local positive effects to the actor,
while related costs are (mainly not directly identifiable) distributed
to others. This effect is referred to a \textbf{\emph{externalities}}
and related to imperfect information.

Joseph Stiglitz, George A. Akerlof and A. Michael Spence jointly received
the Nobel Memorial Prize in Economic Sciences (2001) for their research
in the context of the theory of markets with asymmetric information\cite{Akerlof1970}.
Stiglitz is a distinct critic of the idea of the invisible hand \cite{Smith1759}
applied to economics in the sense that free markets should lead to
efficiency as if guided by unseen forces \cite{Stiglitz2002}. He
points out \cite{Altman2006}:
\begin{quote}
Whenever there are 'externalities' \textendash{} where the actions
of an individual have impacts on others for which they do not pay
or for which they are not compensated \textendash{} markets will not
work well. But recent research has shown that these externalities
are pervasive, whenever there is imperfect information or imperfect
risk markets \textendash{} that is always.

The real debate today is about finding the right balance between the
market and government. Both are needed. They can each complement each
other. This balance will differ from time to time and place to place. 
\end{quote}
Akerlof write in his famous 'lemon markets' paper \cite{Akerlof1970}
about market with \emph{information asymmetry}:
\begin{quote}
It should also be perceived that in these markets social and private
returns differ, and therefore, in some cases, governmental intervention
may increase the welfare of all parties.
\end{quote}
In other words, free markets (without any intervention) will, in general
not lead to efficiency; it is even only under exceptional circumstances
that free markets are efficient. 

Based on these insights, we argue that the formation of the \emph{dEoE
}should not be left over to unregulated markets. It requires dedicated
and continuous methodological support establishing such a dynamically
adapting balance as mentioned by Stiglitz. Not least to prevent a
comparable outcome to Web2.0, where the utility and power aggregated
in the hands of a few quasi platform monopolists, taking advantage
of it to the detriment of smaller competitors and society. One of
the goals of the algorithmic and social mechanisms based self-governance,
in combination with societal governance of \emph{dEOE} endeavors is
to establish functioning market economy. See the following sections
for a more detailed discussion. 

\subsubsection{The detrimental effect of information asymmetry based platforms and
monopolies}

Web2.0 showcases the rise of strong intermediaries in unregulated
complex markets, as already predicted by Akerlof in 1970 \cite{Akerlof1970}.
Due to the pervasive \emph{externalities} and \emph{information asymmetry}
in these markets, there exist a large potential (in the famous automobile
market example of $price/2$) for intermediary merchants:
\begin{quote}
In our picture the important skill of the merchant is identifying
the quality of merchandise; those who can identify used cars in our
example and can guarantee the quality may profit by as much as the
difference between type two traders' buying price and type one traders'
selling price. These people are the merchants.
\end{quote}
In a wider sense, Web2.0 quasi monopolies are built on information
asymmetry, whereas the service provider relates to a merchant ensuring
quality for his customers. The merchant may e.g. deal with quality
search results, goods, networks of like-minded people, machine learning
based services etc. However, the merchant over-proportionately benefits
from the unlocked potential. It is important to note that the merchants
business model, even though he seemingly reduces the perceived uncertainty
of the user, strongly depends on sustaining and extending the information
asymmetry. Hence he is a profiteer and at the same time a preservationist
of market failure, accumulation ever growing advantage in information
and power. As Akerlof states \cite{Akerlof1970}: 
\begin{quote}
... private institutions may arise to take advantage of the potential
increases in welfare which can accrue to all parties. By nature, however,
these institutions are nonatomistic, and therefore concentrations
of power -with ill consequences of their own - can develop.
\end{quote}
Leaving the field to private institutions intensifying information
asymmetry, means allowing draining of ever larger part of the socioeconomic
potential for private profit. In addition, this drainage is cutting
off the rest of the systems participants from the related supermodular
potential. The same holds for (quasi) monopolies in general: they
incorporate a subset of the potential surplus such, that they - as
far as possible - solely benefit, related to over-proportionate aggregation
of surplus and power. This however strongly reduces the total surplus
of the system and hence socioeconomic efficiency.  

In contrast to that, the goals of the democratized Web 3.0 movement
like e.g building trust in trust-less networks, equally accessible
to all participants without intermediaries, preserving privacy or
at least sovereignty over user related information, or more generally
an adequate sharing of effort, benefit, power and responsibilities
is no naive rapture. It is also not about eliminating commercial companies.
On the contrary, it is targeted at establishing a functioning market
economy, characterized by commercial companies not only playing a
role, but having a benefit alongside the society.

\subsubsection{The role of commercial companies, diversity and minority protection }

As argued in the foregoing section, the \emph{dEOE} endeavor is targeted
at a functioning market economy. Commercial companies - alongside
other agents in the network - are contributors and, due to the supermodularity,
benefit from participation. This becomes obvious when considering
commercial companies as providing specialized key expertise generating
surplus in the total system and being rewarded according to their
contribution. Therefore, obviously, investment in competency development
pays off, e.g. for companies. 

As already discussed, \emph{dEOE} is not about naive cooperation,
but \emph{coopetition}. It is important for a functioning market economy
to have a healthy competition, e.g. about providing key expertise.
Diminishing competition and arising monopolies have a detrimental
effect to the total system. Over and above, for complex systems operating
in open contexts, diversity i.a. in the form of competence is important
for antifragility and hence persistence of the system. Therefore,
a competition based diversity on the one hand, balanced by an adequate
protection of minorities form the breeding ground for futures contributors
of key expertise and hence prevailing efficiency.

\subsection{The high level goal\label{subsec:The-high-level}}

Based on the foregoing discussion, we propose the following high level
goal for the \emph{dEOE} - an open ended and adequately decentralized$^{*}$
digital economy of everything ($^{*}$ in the sense of adequate sharing
of effort, benefit, power and responsibilities):
\begin{quote}
The open-ended \emph{dEoE} endeavor lays the foundations for- and
fosters an enduring, non accedence restricted, legal cooperation targeted
at the emergence of an open ended, heterogeneous digital economy by
pooling and appropriately compensating all necessary competencies
and contributions such, that the supermodular potential thereof is
unlocked efficiently, inseparably linked to the collective (no aspect
can over-proportionately be accumulated by some entities, or separated
from the cooperation) and finally leading to the emergence of an adequately
decentralized$^{*}$ \emph{dEoE}. 
\end{quote}

Achieving a true open ended and efficient \emph{dEoE} is, both, a
major challenge and may lead to major benefit and power. It is therefore
neither expectable nor desirable that a few (non-diverse) or even
singular entities successfully initiate and rule the necessary processes
unleashing the supermodular potential, finally leading to utility
and power in the hands of a (few) monopolist(s). 

The necessary efforts, amenities, obligations and responsibilities
therefore each needs to be appropriately shared across an adequate
number of diverse entities. What adequately means depends on the specific
aspect in a specific setting of \emph{context$^{*}$} and \emph{realisation}$^{*}$
(see the \emph{validation triangle$^{*}$} for a detailed discussion).
A basic measure might be derived from the Gini index, possibly extended
by a Shapley value based relation to contribution. 

An adequatly operated \emph{dEOE} will generate a total surplus in
an efficient way. Legality is one of the prerequisites for endurance
and related to acceptance, credibility and hence prevalence. Even
so is non-discriminative accedence. The \emph{dEoE} shall not be a
cooperation of - and for the sake of a few. 

On the other hand, it should not be stoppable by individual nation
states, however needs to be in compliance with international law,
or more generally stated with societal expectations. 

Non-opaqueness of basic \emph{dEOE} modules, competencies and capabilities
is an essential prerequisite to the supermodular potential being unlocked
efficiently (i.e. the ability to combine everything with everything
without access restrictions). However, naive sharing will give rise
to free-riders and fast followers cutting down the benefit of the
innovators and hence it will undermine the goal of the cooperation.
It is rational for agents to behave selfish (they tend to maximize
their own benefit), an appropriate compensation for all necessary
competencies and contributions needs to be established such, that
it is rational for agents to become part of the coopetition based
\emph{dEoE}, complying with the established rules. Maximizing the
benefit from a local perspective (associated to selfishness of agents)
is not evil per se, but the reflection of a necessary contribution
to efficiency of the total system. It needs to be counterbalanced
such, that entity-local maximization leads to maximization of globally
desirable outcomes (i.e. socio-economic optimal results). This compensation
also addresses the free-rider issue. Counterbalancing is well accepted;
see e.g. the discussion in section \ref{subsec:Prerequisites}. 

Complex \emph{coopetitive} systems formed in the context of selfish
entities need to be coalition proof to prevent any coalition of a
few to take over and over-proportionately benefit from the generated
surplus and hence undermining the goal of efficient maximisation of
total surplus, based on adequate decentralisation of powers. Governance,
opinion - \& decision making, advancement of the framework etc. is
part of the necessary competencies. 

Globally desirable outcomes could, in principle, be welfare maximization,
social welfare maximization, Pareto optimal growth (of e.g. capabilities
and benefits) etc. All of these have pros- and cons. The precise combination
thereof will depend on the concrete \emph{setting$^{*}$} ( see the
\emph{validation triangle$^{*}$} for a detailed discussion). Defining
the metrics is part of the sys$^{2}$val based design process. We
therefore, on this level, stick with the more abstract formulation
'such that the super modular potential ... is unlocked efficiently'. 

Re-investment of benefits in indirectly beneficial activities, e.g.
exploration instead of exploitation, support of diversity and protection
of minorities contribute to the overall efficiency and goal in the
long run. For example exploration under-performs with respect to short-term
benefit compared to exploitation; however it improves capability and
therefore the achievable benefit in the mid to long run. Maintaining
diversity (e.g. of \emph{realizations}$^{*}$) might locally appear
inefficient, but will be key to be prepared for \emph{context}$^{*}$
/\emph{ purpose$^{*}$} shifts in the environment. 

'The supermodular potential is inseparably linked to the collective'
relates to the fact that if any aspect (benefit, power, etc.) might
get over-proportionately accumulated by some entities, or separated
from the cooperation, the potential for manipulation of or even dominance
over the cooperation grows and hence the overall goal of achieveing
a functioning economic market, leading to efficient total surplus
maximization is undermined. 

\subsection{Unacceptable losses and hazardous system states}

Based on the foregoing discussion, we propose the following set of
\emph{unacceptable losses} and \emph{hazardous system states} for
the \emph{dEoE}. Note that this is an intentionally condensed set,
from which the aspects discussed in the foregoing sections can be
derived. A more specific and concrete elaboration for the various
levels of the design and operation process is part of the sys$^{2}$val
based approach. 

\subsubsection*{High level unacceptable losses\label{subsec:Unacceptable-losses}}
\begin{enumerate}
\item \emph{dEoE} capability development (and hence benefit to all contributors)
stagnating or falling short. 
\item Potential contributors are not willing to get part of the cooperation
or are leaving the cooperation (to a greater extent). 
\item Individuals or coalitions of a few are dominating or over-proportionately
influencing the opinion - \& decision making, advancement of the framework
of the cooperation or are over-proportionately taking advantage of
the utility for their own sake. 
\item Cooperation repeatedly or massively limited by legal regulations.
Possibly finally leading to (quasi) shut down of the cooperation and
hence the \emph{dEoE}. 
\end{enumerate}

\subsubsection*{High level hazardous system states\label{subsec:Hazardous-system-states}}
\begin{enumerate}
\item The necessary efforts, amenities, obligations and responsibilities
are not appropriately shared across an adequate number of diverse
entities. 
\item Contributor can not participate in proportion to his contributions.
\item Contributor does not (actively) participate in proportion to his contributions.
\item Over-proportionate accumulation of any aspect by individuals or coalitions
of a few. 
\item Contributor network and capability growing slow, stagnating or even
shrinking. 
\item Cooperation perceived as suspicious to potential or active contributors
(e.g. perceived as not credible, suspect to inadequate sharing of
any aspects, discriminating, ...)
\item Cooperation and \emph{dEoE} suspicious to the general public or authorities,
probably leading to legal regulation. 
\item Inadequate diversity in any aspect / type of contribution({*}1) 
\end{enumerate}
An important aspect regarding the establishment of Web3.0, and especially
crypto-based endeavors is the adequate wage of the early contributors
(innovators, investors etc.), on the one hand, while not making permanent
a special role for them, leading to over-proportionate influence and
benefit. The latter leading to scepticism of potential subsequent
contributors and hence under-performance or even stagnation related
to a situation of being \textbf{\emph{locked-in}} in a sub-optimal
system state (e.g. every party benefits much less then possible compared
to a \emph{coopetition} style cooperation). Every party so to say
fears to reinforce the powers of the competitors, in the long term
leading to a marginalization of herself. These aspects are multiply
addressed by the above listed hazardous system states. 

\section{Conclusion}

Based on a detailed discussion of the subject, we argue that the upcoming
digital economy, \emph{dEoE, }corresponds to a \emph{complex system
operated in open contexts}, for which a sophisticated design approach
and a set of dedicated measures are inevitable, in order to be able
to achieve a valid design and operation. In addition, we emphasize
that the high level guiding principles necessary for a valid and effective
\emph{dEoE }align well with the ideals of the democratized Web3.0
movement, and are no naive rapture, but rather - adequate adoption
provided - lead to a functioning market economy characterized by socioeconomic
efficiency. Over and above, we argue that in this kind of \emph{dEoE,}
commercial companies not only play a role, but benefit alongside the
society. Leaving the field to unregulated markets and private institutions
would result in intensifying information asymmetry,as was the case
for Web2.0, and should therefore be prevented.

On first sight, the necessary approach for the formation of a \emph{valid}
\emph{dEoE} endeavor seems complex. However, there is a huge state
of the art e.g. in human societal organization, self governance of
open source projects etc., from which can be drawn. It is therefore
not necessary to reinvent all aspects anew, but to combine understood
and established elements based on system view centric based approaches
for open context systems, such as e.g. the sys$^{2}$val scheme.

As usual with complex systems operated in open contexts, we can not
a priori design a perfectly valid solution right from the start. Therefore,
it is important to establish the processes and measures allowing for
an \emph{antifragile} evolution of the system, based on an \emph{ongoing
holistic cycle$^{*}$}. Sys$^{2}$val provides the basic principles
and measures, allowing to establish a lightweight \emph{meta governance}.
Together with - still to be be widely discussed and accepted - \emph{high
level guiding principles}, this \emph{meta governance} enables to
approach the \emph{dEoE} endeavor on a sound basis. 

\section{References}

\selectlanguage{american}%
\bibliographystyle{myunsrt}
\bibliography{1C__pod2si_docu_lyx_paperGoalsNConstitution20190819_bib_2019_Episteme,2C__pod2si_docu_lyx_paperGoalsNConstitution20190819_bib_20100901_general}
\selectlanguage{english}%

\end{document}